\newcommand{\be}{\begin{equation}}
\newcommand{\ee}{\end{equation}}
\newcommand{\bea}{\begin{eqnarray}}
\newcommand{\eea}{\end{eqnarray}}
\newcommand{\nn}{\nonumber}
\newcommand{\p}[1]{(\ref{#1})}
\documentstyle[12pt,cite]{article}

\begin{document}
\renewcommand{\thefootnote}{\fnsymbol{footnote}}
\thispagestyle{empty}
\rightline{LNF-95/048 (P)}
\rightline{JINR E2-95-376}
\vspace{1cm}
\begin{center}
{\bf NULL FIELDS REALIZATIONS OF $W_3$ FROM
$W(sl(4),sl(3))$ AND $W(sl(3|1),sl(3))$ ALGEBRAS} \vspace{1.5cm} \\
S. Bellucci${}^a$\footnote{E-mail: bellucci@lnf.infn.it},
S. Krivonos${}^{b}$\footnote{E-mail: krivonos@thsun1.jinr.dubna.su}
and A. Sorin${}^{b}$\footnote{E-mail: sorin@thsun1.jinr.dubna.su}
                    \vspace{1cm} \\
${}^a$INFN-Laboratori Nazionali di Frascati, P.O.Box 13 I-00044 Frascati,
            Italy \\
${}^b$Bogoliubov Laboratory of Theoretical Physics, JINR, Dubna, Russia
\vspace{2.5cm} \\
{\bf Abstract}
\end{center}
We consider the nonlinear algebras $W(sl(4),sl(3))$ and $W(sl(3|1),sl(3))$
and find their realizations in terms of currents spanning conformal
linearizing algebras. The specific structure
of these algebras, allows us to construct realizations
modulo null fields of the $W_3$ algebra that lies in the cosets
$W(sl(4),sl(3))/u(1)$ and $W(sl(3|1),sl(3))/u(1)$.
Such realizations exist for the following values of the
$W_3$ algebra central charge: $c_W=-30,-40/7,-98/5,-2$. The first two
values are listed for the first time, whereas for the remaining values we
get the new realizations in terms of an
arbitrary stress tensor and $u(1)\times sl(2)$ affine currents.
\vfill
\begin{center}
September 1995
\end{center}
\setcounter{page}0
\renewcommand{\thefootnote}{\arabic{footnote}}
\setcounter{footnote}0
\newpage
\section{Introduction}
Nonlinear $W$ algebras have been extensively considered from
different points of view for the last years (see e.g. \cite{a1}).
Despite many impressive results the question of constructing
realizations is still open even for the simplest $W_3$ algebra.
The well known tensoring procedure for constructing the new realizations
starting from the known ones, which works perfectly in the case of linear
algebras,
cannot be applied to $W$ algebras due to their intrinsic nonlinearity.
That is why each new realization gives us a deeper understanding of
the structure of nonlinear algebras.

One of the possible ways to construct the realizations of $W$ algebras
is the conformal linearization procedure \cite{{sk1},{sk2},{sk3}}.
The main idea of this
approach is to embed nonlinear $W$ algebra as a subalgebra
in some linear conformal algebra $W^{lin}$. Once this is done,
then each realization
of the linear algebra $W^{lin}$ gives rise to a realization of $W$.
Recent results in this approach \cite{{sk3},{rm}} show that fortunately
a wide class of nonlinear algebras admits conformal linearization and
therefore the realizations of these algebras can be constructed
systematically.

However the problem of constructing the realizations of nonlinear
algebras is far from its complete solution, due to the existence of the so
called realizations modulo null fields \cite{{a2},{a4},{a3},{a7},{a5},{a6}}.
The simplest example is
given by the $W_3$ algebra, where we allow for the spin 4 null operator
$U_4(z)$ in the operator product of two spin 3 generators
$W(z)$:\footnote{The currents in the
r.h.s. of the OPEs are evaluated at the point $z_2$ and
$z_{12}=z_1-z_2$ with normal ordering understood for the products of
currents.}
\bea
W(z_1)W(z_2) & = &  \frac{c_W}{3z_{12}^6}+
     \frac{2T}{z_{12}^4}+
       \frac{T'}{z_{12}^3}+
         \left[ 2 U_4+\frac{3}{10}T''+\frac{32}{22+5c_W}
            \Lambda\right]
\frac{1}{z_{12}^2} + \nn \\
    & &
   \left[ U_4'+\frac{1}{15}T'''+\frac{16}{22+5c_W}
        \Lambda'\right]\frac{1}{z_{12}} \; ,\label{null}
\end{eqnarray}
with $\Lambda = (TT) -\frac{3}{10}T''$. For $U_4(z)$ to be a null operator
we must require that there be no central term in the OPE of $U_4(z)$ with
itself, so that $<U_4U_4>=0$.
The OPE \p{null} together with the standard OPEs with the Virasoro
stress tensor $T(z)$ are not exactly the $W_3$ OPEs. Nevertheless $U_4(z)$,
being a null operator, can only generate null fields in its OPEs.
Therefore the set of null currents generated in a closed algebra is an
ideal and can be consistently set to zero leaving
us with the realization of $W_3$.

It is clear that the construction of the realizations modulo null fields
is a rather complicated task and can be naturally divided in two parts.

Firstly, because the OPE \p{null} is a subset of the OPEs of some
algebra $\cal W$, larger than $W_3$, we need to know this algebra,
together with the condition for its contraction to $W_3$ (that is, the
spectrum of the central charge that makes the equation $<U_4U_4>=0$
satisfied).
A first attempt to classify the possible $\cal W$ algebras which
admit a contraction to $W_N$ is made in \cite{a6}, where a conjecture for
the spectrum of central charges corresponding to the given contraction
is proposed.

Secondly, we need to construct the realizations of the algebra $\cal W$
for the specific values of the central charge that allow the contraction
of $\cal W$ to $W_3$. Up to now,
such realizations of $W_3$  have been constructed for
the following central charges: $c_W=-2,-114/7$ \cite{a3}, $c_W=4/5$ \cite{a4},
$c_W=-10,-2,4/5$ \cite{a5}, $c_W=-98/5,-10,-2,3/4$ \cite{a6},
$c_W=-2,4/5$ \cite{a7}.

The aim of this Letter is to construct  new realizations of the $W_3$
algebra modulo null fields starting from the simplest nonlinear
(super)algebras $W(sl(4),sl(3))$ and $W(sl(3|1),sl(3))$ which possess
the following properties:
\begin{itemize}
\item the algebras include the OPE \p{null},
\item the algebras can be linearized.
\end{itemize}
The linearizing algebras are useful for the construction of the realizations,
while the presence of the OPE \p{null} guarantees the existence of the
contraction to the $W_3$ algebra.
New realizations exist for the following values of the $W_3$ central
charge: $c_W=-40/7,-2,-30$ ($W(sl(4),sl(3))$)\footnote{Only two of these
values $c_W=-40/7,-2$ coincide with the conjecture in \cite{a6}.},
$c_W=-2,-98/5$ ($W(sl(3|1),sl(3))$). We would like to stress here
that even for the previously known
values of the central charge the
constructed realizations of $W_3$ modulo null fields are new.

\setcounter{equation}0
\section{$W(sl(4),sl(3))$ and $W(sl(3|1),sl(3))$ algebras}

In this Section we present the explicit structure of
$W(sl(4),sl(3))$ and $W(sl(3|1),sl(3))$ algebras in the quantum case.

The $W(sl(4),sl(3))$ algebra can be constructed, as hinted by its very
name, by considering the principal embedding of the $sl(2)$ algebra into
the $sl(3)$ subalgebra of the $sl(4)$ \cite{{a8},{a1}}.
The  $W(sl(4),sl(3))$ algebra exists for a generic central charge $c$.
Its currents obey the following OPEs:

\begin{eqnarray}
T(z_1)T(z_2) & = &  \frac{2c}{z_{12}^4}+
        \frac{2T}{z_{12}^2}+\frac{T'}{z_{12}} , \quad
T(z_1)U(z_2)  =   \frac{U}{z_{12}^2}+\frac{U'}{z_{12}}  ,\nn  \\
T(z_1)G(z_2) & = &  \frac{2G}{z_{12}^2}+\frac{G'}{z_{12}}  , \quad
T(z_1){\overline G}(z_2)  =   \frac{2{\overline G}}{z_{12}^2}+
                           \frac{{\overline G}'}{z_{12}}  ,\quad
T(z_1)W(z_2)  =   \frac{3W}{z_{12}^2}+\frac{W'}{z_{12}}  ,\nn  \\
U(z_1)U(z_2) & = &  -\frac{{c_1}/2}{z_{12}^2}  ,  \quad
U(z_1)G(z_2)  =   -\frac{2G}{z_{12}} ,  \quad
U(z_1){\overline G}(z_2) =  \frac{2{\overline G}}{z_{12}} \quad ,\nn  \\
G(z_1){\overline G}(z_2) & = &  \frac{{c_2}/3}{z_{12}^4}+
   \frac{a_1 U}{z_{12}^3}+\left[ \frac{1}{3}T +a_2 (UU)+a_3 U'\right]
   \frac{1}{z_{12}^2}  \nn \\
     & &
   +\left[ W+a_4 (TU)+a_5 (U'U)+\frac{1}{6} T'+a_6 U''+a_{7}(UUU)\right]
   \frac{1}{z_{12}}  \quad ,   \nn \\
G(z_1)W(z_2) & = &  \frac{b_1 G}{z_{12}^3}+
         \left[ b_2 (UG)+b_3 G'\right] \frac{1}{z_{12}^2} \nn \\
   & & +\left[ b_4 (TG)+b_5 (UUG)+b_6 (U'G)+b_7 (UG')+b_8 G''\right]
    \frac{1}{z_{12}}  \quad , \nn  \\
{\overline G}(z_1)W(z_2) & = &  -\frac{b_1{\overline G}}{z_{12}^3}+
\left[ b_2(U{\overline G})-b_3{\overline G}'\right] \frac{1}{z_{12}^2}\nn\\
     & &
     -\left[ b_4 (T{\overline G})+b_5 (UU{\overline G})-b_6
     (U'{\overline G})-b_7 (U{\overline G}')+b_8 {\overline G}''\right]
     \frac{1}{z_{12}}    \quad , \nn \\
W(z_1)W(z_2) & = &  A\left\{ \frac{c_W}{3z_{12}^6}+
     \frac{2{\tilde T}}{z_{12}^4}+
       \frac{{\tilde T}{}'}{z_{12}^3}+
         \left[ 2 U_4+\frac{3}{10}{\tilde T}''+\frac{32}{22+5c_W}
            \Lambda\right]
\frac{1}{z_{12}^2} \right.  \nn \\
    & &
   +\left.\left[ U_4'+\frac{1}{15}{\tilde T}'''+\frac{16}{22+5c_W}
        \Lambda'\right]\frac{1}{z_{12}} \right\} ,\label{Wnew}
\end{eqnarray}
where
\bea
{\tilde T} & = & T+\frac{1}{c_1}(UU), \label{newt}\\
\Lambda   & = & {\tilde T}{\tilde T} -\frac{3}{10}{\tilde T}'' ,
                              \label{lambda} \\
U_4 & = & d_1 (G{\overline G})+d_2W'+d_3(WU)+
        d_4 (T T)+d_5 (TUU)+
        d_6 (T'U)+d_7 (TU') \nn \\
    & &  + d_8 T''+  d_9 (UUUU)+d_{10}(U'UU)+
        d_{11} (U'U')+d_{12}(U''U)+d_{13} U'''.  \label{u4}
\eea
and the values of all coefficients are given in the Table. We introduce
the coefficient $A$ in the r.h.s. of the $W(z_1)W(z_2)$ OPE in \p{Wnew} for
convenience. For sure, it can be set to one by rescaling the current
$W(z)$, at the cost, however, of further complicating the coefficients in the
Table.

Let us note that the spin 4 current $U_4(z)$ is defined to be primary
with respect to the new stress tensor ${\tilde T}(z)$\footnote{Due to
the regular OPE of the spin 3 current $W(z)$ with $U(z)$, $W(z)$ is still
a primary current, also with respect to $\tilde T$.}
\be
{\tilde T}(z_1){\tilde T}(z_2) =   \frac{c_W}{2z_{12}^4}+
        \frac{2{\tilde T}}{z_{12}^2}+\frac{{\tilde T}'}{z_{12}} , \quad
{\tilde T}(z_1)U_4(z_2)  =   \frac{4U_4}{z_{12}^2}+\frac{U_4'}{z_{12}}
\ee
and both $\tilde T$ and $U_4$ have regular OPEs with the $u(1)$ current $U(z)$
\be
U(z_1){\tilde T}(z_2) = U(z_1)U_4(z_2) = \mbox{regular}.  \label{coset}
\ee
Thus, the currents ${\tilde T}(z),W(z)$ and $U_4(z)$ belong to the
coset $W(sl(4),sl(3))/u(1)$.

We would like also to stress that there is no possibility to redefine
the currents of $W(sl(4),sl(3))$, in order to avoid the appearance of
the $U_4(z)$ current in the r.h.s. of the
OPE $W(z_1)W(z_2)$. Therefore, the
$W(sl(4),sl(3))$ algebra does not contain the $W_3$ one as a subalgebra.

The $W(sl(3|1),sl(3))$ superalgebra contains currents with the same
conformal spins as the $W(sl(4),sl(3))$ ones: the Virasoro stress
tensor $T(z)$, a bosonic spin 1 current $U(z)$, a
doublet of fermionic spin 2 currents $G(z)$
and ${\overline G}(z)$, and a bosonic spin 3 current $W(z)$. So the only
differences in the contents of $W(sl(3|1),sl(3))$ and
$W(sl(4),sl(3))$ algebras is the statistic of the spin 2
currents doublet. This is why the OPEs for the $W(sl(3|1),sl(3))$
algebra can be written in the form \p{Wnew} with the same definitions
for the composite currents \p{newt}-\p{u4} and the coefficients
given in the Table.

{\small
\begin{center}
\begin{tabular}{lcc||lcc}\hline
Coeff.  & W(sl(4),sl(3)) &  W(sl(3$|$1),sl(3)) &
Coeff.  & W(sl(4),sl(3)) &  W(sl(3$|$1),sl(3)) \\ \hline
$c$     & $\frac{c_1(4c_1+13)}{4(c_1-8)}$ &
          $-\frac{c_1(2c_1+7)}{4(c_1+8)}$ &
$b_1$   & $\frac{2(c_1+2)(5c_1-16)}{9c_1^2}$  &
           $\frac{2(c_1-8)(c_1+2)(5c_1+16)}{9c_1^2(c_1+8)}$ \\
$c_2$     & $\frac{c_1(c_1+1)}{(c_1-8)}$ &
          $-\frac{c_1(c_1+2)}{2(c_1+8)}$ &
$b_2$   & $\frac{2(5c_1-16)}{3c_1^2}$  &
           $\frac{2(c_1-8)(5c_1+16)}{3c_1^2(c_1+8)}$ \\
$c_W$     & $\frac{4(c_1+1)(c_1+2)}{c_1-8}$ &
          $-\frac{2(c_1+2)^2}{c_1+8}$ &
$b_3$   & $\frac{(c_1+8)(5c_1-16)}{18c_1^2}$  &
           $\frac{(c_1-8)(5c_1+16)}{18c_1^2}$ \\
$a_1$     & $\frac{4(c_1+1)}{3(c_1-8)}$ &
          $-\frac{2(c_1+2)}{3(c_1+8)}$ &
$b_4$   & $\frac{4(c_1-8)}{3c_1(c_1+4)}$  &
           $-\frac{8}{3c_1}$ \\
$a_2$     & $\frac{3}{c_1-8}$ &
          $-\frac{1}{c_1+8}$ &
$b_5$   & $\frac{4(5c_1-16)}{3c_1^2(c_1+4)}$  &
           $\frac{8(c_1-16)}{3c_1^2(c_1+8)}$ \\
$a_3$     & $\frac{2(c_1+1)}{3(c_1-8)}$ &
          $-\frac{c_1+2}{3(c_1+8)}$ &
$b_6$   & $\frac{2}{c_1+4}$  &
           $\frac{2}{c_1+8}$ \\
$a_4$     & $\frac{4}{3c_1}$ &
          $\frac{4}{3c_1}$ &
$b_7$   & $\frac{2(c_1-4)(c_1+16)}{3c_1^2(c_1+4)}$  &
           $\frac{2(c_1-16)}{3c_1^2}$ \\
$a_5$     & $\frac{3}{c_1-8}$ &
          $-\frac{1}{c_1+8}$ &
$b_8$   & $\frac{c_1^3+4c_1^2+80c_1-256}{18c_1^2(c_1+4)}$  &
           $\frac{(c_1-4)(c_1+16)}{18c_1^2}$ \\
$a_6$     & $\frac{2(c_1^2-2c_1+24)}{9c_1(c_1-8)}$ &
          $-\frac{c_1^2+8c_1+48}{9c_1(c_1+8)}$ &
$A$   & $-\frac{5c_1-16}{18c_1}$  &
           $-\frac{(c_1-8)(5c_1+16)}{18c_1(c_1+8)}$ \\
$a_7$     & $\frac{4(11c_1-16)}{9c_1^2(c_1-8)}$ &
          $-\frac{4(c_1-16)}{9c_1^2(c_1+8)}$ &  &
          & \\
\hline\hline
Coeff.  & \multicolumn{2}{c}{W(sl(4),sl(3))} &
\multicolumn{3}{c}{  W(sl(3$|$1),sl(3))} \\ \hline
$d_1$ & \multicolumn{2}{c}{$-\frac{72c_1}{(5c_1-16)(c_1+4)}$} &
\multicolumn{3}{c}{$-\frac{144c_1}{(5c_1+16)(c_1-8)}$  } \\
$d_2$ & \multicolumn{2}{c}{$\frac{36c_1}{(5c_1-16)(c_1+4)}$} &
\multicolumn{3}{c}{$\frac{72c_1}{(5c_1+16)(c_1-8)}$  } \\
$d_3$ & \multicolumn{2}{c}{$\frac{288}{(5c_1-16)(c_1+4)}$} &
\multicolumn{3}{c}{$\frac{576}{(5c_1+16)(c_1-8)}$  } \\
$d_4$ & \multicolumn{2}{c}{$-\frac{12(c_1-8)(11c_1+20)}
             {(5c_1-16)(c_1+4)(10c_1^2+41c_1-68)}$} &
\multicolumn{3}{c}{$\frac{24(c_1+8)(11c_1+20)}
            {(5c_1+16)(c_1-8)(5c_1^2+9c_1-68)}$  } \\
$d_5$ & \multicolumn{2}{c}{$-\frac{12(91c_1^2+260c_1-704)}
             {(5c_1-16)(c_1+4)(10c_1^2+41c_1-68)}$} &
\multicolumn{3}{c}{$\frac{48(29c_1^2-36c_1-704)}
            {(5c_1+16)(c_1-8)(5c_1^2+9c_1-68)}$  } \\
$d_6=d_7$ & \multicolumn{2}{c}{$\frac{48}
             {(5c_1-16)(c_1+4)}$} &
\multicolumn{3}{c}{$\frac{96}
            {(5c_1+16)(c_1-8)}$  } \\
$d_8$ & \multicolumn{2}{c}{$\frac{36(c_1^2-c_1+4)}
             {(5c_1-16)(10c_1^2+41c_1-68)}$} &
\multicolumn{3}{c}{$\frac{36(c_1+4)(c_1^2+8)}
            {(5c_1+16)(c_1-8)(5c_1^2+9c_1-68)}$  } \\
$d_9$ & \multicolumn{2}{c}{$\frac{4(1153c_1^3+924c_1^2-19584c_1+25600)}
             {(c_1-8)c_1^2(5c_1-16)(c_1+4)(10c_1^2+41c_1-68)}$} &
\multicolumn{3}{c}{$\frac{8(47c_1^3+1156c_1^2-1280c_1-25600)}
            {(5c_1+16)c_1^2(c_1-8)(c_1+8)(5c_1^2+9c_1-68)}$  } \\
$d_{10}$ & \multicolumn{2}{c}{$\frac{48(11c_1-16)}
             {(c_1-8)c_1(5c_1-16)(c_1+4)}$} &
\multicolumn{3}{c}{$\frac{-96(c_1-16)}
            {(5c_1+16)c_1(c_1-8)(c_1+8)}$  } \\
$d_{11}$ & \multicolumn{2}{c}{$\frac{24(c_1-2)(23c_1^3+133c_1^2+140c_1+192)}
             {(c_1-8)c_1(5c_1-16)(c_1+4)(10c_1^2+41c_1-68)}$} &
\multicolumn{3}{c}{$-\frac{24(7c_1^4+2c_1^3-220c_1^2-560c_1-768)}
            {(5c_1+16)c_1(c_1-8)(c_1+8)(5c_1^2+9c_1-68)}$  } \\
$d_{12}$ &
\multicolumn{2}{c}{$\frac{4(178c_1^4+213c_1^3+1476c_1^2+17056c_1-32256)}
             {(c_1-8)c_1(5c_1-16)(c_1+4)(10c_1^2+41c_1-68)}$} &
\multicolumn{3}{c}{$-\frac{8(31c_1^4+251c_1^3+1004c_1^2-4832c_1-32256)}
            {(5c_1+16)c_1(c_1-8)(c_1+8)(5c_1^2+9c_1-68)}$  } \\
$d_{13}$ & \multicolumn{2}{c}{$\frac{4(c_1^2-5c_1+48)}
             {(5c_1-16)(c_1-8)(c_1+4)}$} &
\multicolumn{3}{c}{$\frac{4(c_1^2+14c_1+96)}
            {(5c_1+16)(c_1-8)(c_1+8)}$  } \\
\hline
\end{tabular}\vspace{0.5cm}\\
Table.
\end{center}
}

\setcounter{equation}0
\section{Contractions of $W(sl(3|1),sl(3))$ and $W(sl(4),sl(3))$
to $W_3$ algebra}
As stated in the Introduction, our purpose is building new realizations of
$W_3$ modulo null fields and finding the corresponding values
of the central charge. From the explicit OPEs of the $W(sl(4),sl(3))$ and
$W(sl(3|1),sl(3))$ algebras \p{Wnew} we can see that the currents
${\tilde T}(z)$ \p{newt} and $W(z)$ form a $W_3$ algebra,
modulo the spin 4 current
$U_4(z)$ which is present in the r.h.s. of OPE $W(z_1)W(z_1)$.
This spin 4 current $U_4$ in both cases
is expressed in terms of the basic currents \p{u4}.
Therefore we can
require that the current $U_4(z)$ be a null operator, i.e. $< U_4U_4 > =0$.
For this, all we need to ask is the vanishing of the central term in the
$U_4(z_1)U_4(z_2)$ OPE.
The corresponding equation is satisfied only for some special values of the
central charge.

Next, we give the results for both  $W(sl(4),sl(3))$ and
$W(sl(3|1),sl(3))$.

\subsection{$W(sl(4),sl(3))$ case}
Here we list for completeness the vacuum expectation values for
the currents ${\tilde T}, W, U_4$ contained in
the coset $W(sl(4),sl(3))/u(1)$
\bea
< {\tilde T}{\tilde T} > & = & \frac{ 2(c_1+1)(c_1+2)}{(c_1-8)} , \\
< WW > & = & -\frac{ 2(c_1+1)(c_1+2)(5c_1-16)}{27c_1(c_1-8)} , \\
< U_4U_4 > & = & \frac{ 576(c_1-4)(c_1+1)(c_1+2)(c_1+6)(2c_1-1)}
    {(c_1-8)(c_1+4)(5c_1-16)(10c_1^2+41c_1-68)} .
\eea
Thus for the values $c_1=-6,\frac{1}{2},4$ which correspond to the
following values of the $W_3$ central charge $c_W$:
\bea
c_1 = -6 & \Rightarrow & c_W=-\frac{40}{7} , \\
c_1 = \frac{1}{2} & \Rightarrow & c_W=-2 , \\
c_1 = 4 & \Rightarrow & c_W=-30 ,
\eea
the spin 4 current $U_4$ becomes a null operator. All other poles
and zeros of the vacuum expectation value $<U_4U_4>$ provide us with
further contractions of the algebra, where the
spin 3 current $W(z)$ and even the stress tensor
${\tilde T}(z)$ become null operators.

Let us note that only the first two values of the $W_3$ central charge
$c_W=-40/7,-2$
follow from the conjecture of \cite{a6}. Therefore the spectrum
of central charges for the contraction of the
$W(sl(4),sl(3))$ algebra to $W_3$
proposed in \cite{a6} is not exhaustive.

\subsection{$W(sl(3|1),sl(3))$ case}
The same calculations for the $W(sl(3|1),sl(3))$ superalgebra give
the following results:
\bea
< {\tilde T}{\tilde T} > & = & \frac{ -(c_1+2)^2}{(c_1+8)} , \\
< WW > & = & \frac{ (c_1-8)(c_1+2)^2(5c_1+16)}{27c_1(c_1+8)^2} , \\
< U_4U_4 > & = & -\frac{576 (c_1-12)(c_1-1)(c_1+2)^2(c_1+4)^2}
    {(c_1-8)(c_1+8)^2(5c_1+16)(5c_1^2+9c_1-68)} .
\eea
So, the current $U_4$ is a null operator for the following values:
\bea
c_1 = -4 & \Rightarrow & c_W=-2 , \\
c_1 = 1 & \Rightarrow & c_W=-2 , \\
c_1 = 12 & \Rightarrow & c_W=-\frac{98}{5} .
\eea

This concludes the determination of the central charges spectrum
for both $W(sl(4),sl(3))$ and $W(sl(3|1),sl(3))$, when these algebras
are contracted to the $W_3$ one.
In order to complete our task, we need to construct the realization
of $W_3$ modulo null fields explicitly. In the next Section we will show
that a straightforward way to construct such realizations comes from
the conformal linearization procedure
\cite{{sk1},{sk2},{sk3}} applied to the algebras under
consideration.

\setcounter{equation}0
\section{Null fields realizations of $W_3$ algebra}
One of the most important questions when considering nonlinear
$W$ algebras is the construction of their realizations in terms of
free fields or affine currents. The lack of a tensoring procedure
for constructing new realizations starting from the known ones makes the
task of finding new realizations of the nonlinear algebras rather
difficult. Moreover, in many cases it is unclear which set of
currents we need to use, in order to construct such realizations.

A solution to this problem was proposed in
\cite{{sk1},{sk2},{sk3}}. The idea of this approach is to embed the given
nonlinear algebra into a {\it conformal linear} one which contains the former
nonlinear algebra as a subalgebra. Of course, such linearizing algebra
contains more currents, in comparison with the nonlinear one.
However, after performing such a linearization, the
question of constructing realizations becomes almost trivial, because
any realization of the linear algebra gives rise to a
realization of the corresponding nonlinear one.

\subsection{Linearization of $W(sl(4),sl(3))$ algebra}
Fortunately, for $W(sl(4),sl(3))$ such a linear algebra
which contains it as a subalgebra is known \cite{sk3}. It contains
in the primary basis, besides the Virasoro stress tensor ${\cal T}(z)$,
four currents with spin 1, i.e.
${\cal U}(z),{\cal J}_+(z),{\cal J}_3(z),{\cal J}_-(z)$ forming the
$u(1)\times sl(2)$ affine algebra, and two additional currents
${\cal G}^+_1(z)$ and ${\cal G}^+_2(z)$ with the unusual spin
$\frac{3K+4}{2K}$,
where $K$ is the level of the $u(1)$ affine
algebra.\footnote{The currents
of the $sl(2)$ affine algebra are related to those in the
paper \cite{sk3} as follows: ${\cal J}_3=J_1^1,{\cal J}_+=J_1^2,
{\cal J}_-=-J^1_2$.} The
complete list of OPEs for this linear algebra reads as follows:
\begin{eqnarray}
{\cal T}(z_1){\cal T}(z_2) & = &
    \frac{(3-2K)(3K-4)}{2Kz_{12}^4}+
    \frac{2{\cal T}}{z_{12}^2}+
    \frac{{\cal T}'}{z_{12}}, \nn  \\
{\cal T}(z_1){\cal U}(z_2) & = &
    \frac{{\cal U}}{z_{12}^2}+\frac{{\cal U}'}{z_{12}}, \quad
{\cal T}(z_1){\cal J}_+ (z_2)  =
    \frac{{\cal J}_+}{z_{12}^2}+\frac{{{\cal J}_+}^{'}}{z_{12}}, \nn \\
{\cal T}(z_1){\cal J}_- (z_2) & = &
    \frac{{\cal J}_-}{z_{12}^2}+\frac{{{\cal J}_-}^{'}}{z_{12}}, \quad
{\cal T}(z_1){\cal J}_3 (z_2)  =
    \frac{{\cal J}_3}{z_{12}^2}+\frac{{{\cal J}_3}^{'}}{z_{12}} \quad ,\nn \\
{\cal T}(z_1){\cal G}^+_1 (z_2) & = &
    \frac{(3K+4){\cal G}^+_1}{2Kz_{12}^2}+\frac{{{\cal G}^+_1}^{'}}{z_{12}},
\quad
{\cal T}(z_1){\cal G}^+_2 (z_2)  =
    \frac{(3K+4){\cal G}^+_2}{2Kz_{12}^2}+\frac{{{\cal G}^+_2}^{'}}{z_{12}},\nn
\\
{\cal U}(z_1){\cal U}(z_2) & = &  \frac{K}{z_{12}^2}, \quad
{\cal U}(z_1){\cal G}^+_1 (z_2)  =   \frac{{\cal G}^+_1}{z_{12}},  \quad
{\cal U}(z_1){\cal G}^+_2 (z_2)  =   \frac{{\cal G}^+_2}{z_{12}},   \nn \\
{\cal J}_+ (z_1){\cal J}_- (z_2) & = &
    \frac{2-K}{z_{12}^2}-\frac{2{\cal J}_3}{z_{12}}, \quad
{\cal J}_+ (z_1){\cal J}_3 (z_2)  =    -\frac{{\cal J}_+}{z_{12}}, \quad
{\cal J}_+ (z_1){\cal G}^+_1 (z_2)  =    -\frac{{\cal G}^+_2}{z_{12}}, \nn \\
{\cal J}_- (z_1){\cal J}_3 (z_2) & = &    \frac{{\cal J}_-}{z_{12}}, \quad
{\cal J}_- (z_1){\cal G}^+_2 (z_2)  =     \frac{{\cal G}^+_1}{z_{12}}, \quad
{\cal J}_3 (z_1){\cal J}_3 (z_2)  =   \frac{K-2}{2z_{12}^2},  \nn \\
{\cal J}_3 (z_1){\cal G}^+_1 (z_2) & = &   -\frac{{\cal G}^+_1}{2z_{12}}, \quad
{\cal J}_3 (z_1){\cal G}^+_2 (z_2)  =     \frac{{\cal G}^+_2}{2z_{12}} \quad .
\label{WnewBosLin}
\eea
One may wonder how it could be possible to construct the currents of
$W(sl(4),sl(3))$ from the currents of the linear algebra
\p{WnewBosLin} which possess completely different spins. The answer
becomes clear after defining the new stress tensor $T(z)$
\be
T(z) = {\cal T} +{\cal J}_3'-\frac{K-2}{K} {\cal U}'\quad . \label{tt}
\ee
With respect to this stress tensor the currents of the linear algebra
\p{WnewBosLin} have the following spins: \vspace{0.1cm} \\
\begin{center}
\begin{tabular}{|l|c|c|c|c|c|c|} \hline
Currents & ${\cal U}$ & ${\cal J}_+$ & ${\cal J}_-$ & ${\cal J}_3$
    & ${\cal G}^+_1$ & ${\cal G}^+_2$\\ \hline
Spins    &   1      &   0   &   2   &   1   &    3    &    2    \\ \hline
\end{tabular}
\end{center}
So, with respect to the stress tensor $T(z)$ \p{tt} the currents of the linear
algebra \p{WnewBosLin} possess the spins needed, in order to construct from
them the currents forming the $W(sl(4),sl(3))$ algebra.
Moreover, after imposing the following relation between the free parameter
$c_1$ in $W(sl(4),sl(3))$ and the affine level $K$ in the
linear algebra:
\be
c_1=8-6K \; ,
\ee
the central charge for the stress tensor $T(z)$ \p{tt} coincides with
the central charge of the Virasoro subalgebra of \p{Wnew}. Therefore
we can identify these two stress tensors (that is why we used the same letter
in the definition \p{tt}).

Now it is a matter of straightforward
calculations to find the expressions for the remaining
currents of the nonlinear $W(sl(4),sl(3))$ algebra in terms of the currents
spanning the linear algebra \p{WnewBosLin}

\begin{eqnarray}
U & = & {\cal U} +2{\cal J}_3,\quad {\cal G}  =  {\cal J}_- , \nn \\
{\overline G} & = & {\cal G}^+_2 -\frac{({\cal T}J_+)}{3(K-2)}-
\frac{({\cal J}_+{\cal J}_+{\cal J}_-)}{3K(K-2)}+
\frac{2({\cal J}_+{\cal J}_3{\cal J}_3)}{3K(K-2)}-
\frac{({\cal J}_+{\cal J}_3')}{3(K-2)}+
\frac{2({\cal U}{\cal J}_+{\cal J}_3)}{3K(K-2)}+  \nn \\
& & \frac{({\cal U}{\cal U}{\cal J}_+)}{2K(K-2)}-
\frac{2({\cal U}{\cal J}_+')}{3K}-\frac{2({\cal J}_+'{\cal J}_3)}{3K}+
\frac{(1-K)({\cal U}'{\cal J}_+)}{3K(K-2)}+
\frac{(3-3K+K^2){\cal J}_+''}{3K(K-2)}, \nn  \\
W & = & {\cal G}^+_1 -\frac{({\cal J}_+{\cal J}_-')}{3K}+
\frac{4(5K-4)({\cal J}_3{\cal J}_3{\cal J}_3)}{9(K-2)(3K-4)^2}+
\frac{2({\cal U}{\cal J}_+{\cal J}_-)}{3K(K-2)}-
\frac{8(K^2-5K+4)({\cal U}{\cal J}_3{\cal J}_3)}{3K(K-2)(3K-4)^2}+\nn \\
    & &
\frac{(5K-4)({\cal U}{\cal U}{\cal J}_3)}{3(K-2)(3K-4)^2}+
\frac{(12-11K)({\cal U}{\cal U}{\cal U})}{9K(3K-4)^2}+
\frac{(4-K)({\cal U}{\cal J}_3')}{3K(3K-4)}
    + \frac{2(1-K)({\cal J}_+'{\cal J}_-)}{3K(K-2)}+\nn \\
    & &
\frac{4(K^2-5K+4)({\cal J}_3'{\cal J}_3)}{3K(K-2)(3K-4)}-
\frac{K({\cal U}'{\cal J}_3)}{3(K-2)(3K-4)}+
\frac{(5K-4)({\cal U}'{\cal U})}{6K(3K-4)}+
+\frac{2({\cal T}{\cal U})}{3(3K-4)}\nn \\
    & &
\frac{(3K^3-22K^2+68K-48){\cal J}_3''}{18K(K-2)(3K-4)}-
\frac{(3K+2){\cal U}''}{18(3K-4)}-
\frac{2K({\cal T}{\cal J}_3)}{3(K-2)(3K-4)}
 -\frac{{\cal T}'}{6}. \label{realiz1}
\end{eqnarray}

The expressions above, in spite of their rather complicated
appearence, allow us to construct the realizations of the nonlinear
$W(sl(4),sl(3))$ algebra starting from any given realization
of the linear algebra \p{WnewBosLin}. The simplest realization
corresponds to the case when both the ${\cal G}^+_1$ and ${\cal G}^+_2$
currents
are vanishing (i.e. they are null fields in the algebra \p{WnewBosLin}) and so
we are left with the realizations of the nonlinear $W(sl(4),sl(3))$
algebra in terms of an arbitrary stress
tensor ${\cal T}(z)$, and the $u(1)\times sl(2)$ affine currents
${\cal U}(z),{\cal J}_+(z),{\cal J}_-(z),{\cal J}_3(z)$.

Let us note that the three exceptional points $K=0,4/3,2$,
where the transformations \p{realiz1} become singular,
correspond to the central charges $c_1=8,0,-4$ in the
$W(sl(4),sl(3))$ algebra. At these points some of the coefficients
in the Table are also singular and hence the currents that span the
$W(sl(4),sl(3))$ algebra must be redefined, in order to avoid the singularities
(see e.g. \cite{a3}). So, the appearance of singular terms in the
transformations from the linearizing algebra \p{WnewBosLin} to the nonlinear
$W(sl(4),sl(3))$ one is dictated by the structure relations for the
$W(sl(4),sl(3))$ algebra we start with.

Thus, we succeed in the construction of the realizations \p{realiz1}
for $W(sl(4),sl(3))$. For the following values of the parameter:
$K=7/3, 5/4, 2/3$,
which correspond to the central charge (3.4)-(3.6), the realization
\p{realiz1} is {\it the realization of $W_3$ modulo null fields}.

\subsection{Linearization of $W(sl(3|1),sl(3))$ superalgebra}
The conformal linearizing superalgebra for $W(sl(3|1), sl(3))$
has not been constructed so far.
However the bosonic case we considered
in the previous Section gives us some hints
how such linear conformal superalgebra
can be constructed. It is natural to assume that the linearizing superalgebra
for $W(sl(3|1),sl(3))$ must contain the same spins as its bosonic
counterpart \p{WnewBosLin} with some currents being now fermionic ones.

Without further justifications,
we write down the linearizing superalgebra which
contains the following currents: a Virasoro stress tensor ${\cal T}(z)$,
four currents with spin 1, i.e. the bosonic currents
${\cal U}(z),{\cal J}_3(z)$ and the fermionic ones
${\cal J}_+(z),{\cal J}_-(z)$,
as well as two additional fermionic currents
${\cal G}^+_1(z)$ and ${\cal G}^+_2(z)$ with spin $\frac{3K-32}{2(K-4)}$.
The currents obey the OPEs

\begin{eqnarray}
{\cal T}(z_1){\cal T}(z_2) & = &
    \frac{(K+11)(3K+8)}{10(K-4)z_{12}^4}+
    \frac{2{\cal T}}{z_{12}^2}+
    \frac{{\cal T}'}{z_{12}}, \quad
{\cal T}(z_1){\cal U}(z_2)  =
    \frac{{\cal U}}{z_{12}^2}+\frac{{\cal U}'}{z_{12}}, \nn \\
{\cal T}(z_1){\cal J}_+ (z_2)  & = &
    \frac{{\cal J}_+}{z_{12}^2}+\frac{{{\cal J}_+}^{'}}{z_{12}}, \quad
{\cal T}(z_1){\cal J}_- (z_2)  =
    \frac{{\cal J}_-}{z_{12}^2}+\frac{{{\cal J}_-}^{'}}{z_{12}}, \quad
{\cal T}(z_1){\cal J}_3 (z_2)  =
    \frac{{\cal J}_3}{z_{12}^2}+\frac{{{\cal J}_3}^{'}}{z_{12}}, \nn \\
{\cal T}(z_1){\cal G}^+_1 (z_2) & = &
    \frac{(3K-32){\cal G}^+_1}{2(K-4)z_{12}^2}+
  \frac{{{\cal G}^+_1}^{'}}{z_{12}}, \quad
{\cal T}(z_1){\cal G}^+_2 (z_2)  =
    \frac{(3K-32){\cal G}^+_2}{2(K-4)z_{12}^2}+
 \frac{{{\cal G}^+_2}^{'}}{z_{12}},\nn \\
{\cal U}(z_1){\cal U}(z_2) & = &  \frac{K}{z_{12}^2}, \quad
{\cal U}(z_1){\cal G}^+_1 (z_2)  =   \frac{{\cal G}^+_1}{z_{12}},  \quad
{\cal U}(z_1){\cal G}^+_2 (z_2)  =   \frac{3{\cal G}^+_2}{z_{12}},   \nn \\
{\cal U}(z_1){\cal J}_+ (z_2) & = &  \frac{2{\cal J}_+}{z_{12}},   \quad
{\cal U}(z_1){\cal J}_- (z_2)  =   -\frac{2{\cal J}_-}{z_{12}},   \quad
{\cal U}(z_1){\cal J}_3 (z_2)  =   -\frac{K-4}{5z_{12}^2},   \nn \\
{\cal J}_+ (z_1){\cal J}_- (z_2) & = &
  -\frac{K-4}{10z_{12}^2}+\frac{{\cal J}_3}{z_{12}},  \quad
{\cal J}_+ (z_1){\cal G}^+_1 (z_2)  =    -\frac{{\cal G}^+_2}{z_{12}}, \nn \\
{\cal J}_- (z_1){\cal G}^+_2 (z_2)  & = &    \frac{{\cal G}^+_1}{z_{12}}, \quad
{\cal J}_3 (z_1){\cal G}^+_1 (z_2)  =    -\frac{{\cal G}^+_1}{z_{12}}, \quad
{\cal J}_3 (z_1){\cal G}^+_2 (z_2)  =     \frac{{\cal G}^+_2}{z_{12}} \quad ,
\label{superlin}
\eea
where the free parameter
$c_1$ in $W(sl(3|1),sl(3))$ and the affine level $K$ in the
linearizing superalgebra \p{superlin} are related as follows:
\be
c_1=\frac{8+3K}{5} \; .
\ee

Analogously to the bosonic case, we can express the
$W(sl(3|1),sl(3))$ currents in terms of the currents of the linear
superalgebra \p{superlin}:

\begin{eqnarray}
T & = & {\cal T}+\frac{2(K+1)}{K-4}{\cal J}_3'+\frac{1}{2}{\cal U}', \nn \\
U & = & {\cal U} +{\cal J}_3,\quad G  =  {\cal J}_- , \nn \\
{\overline G} & = & {\cal G}^+_2 +\frac{10({\cal T}{\cal J}_+)}{3(K-4)}-
\frac{75({\cal J}_+{\cal J}_3{\cal J}_3)}{(K-4)^2}-
 \frac{10({\cal J}_+{\cal J}_3')}{3(K-4)^2}+
\frac{50({\cal U}{\cal J}_+{\cal J}_3)}{(K-4)^2}+
\frac{25({\cal U}{\cal U}{\cal J}_+)}{3(K-4)^2}- \nn \\
& &
\frac{10(K+6)({\cal U}{\cal J}_+')}{3(K-4)^2}-
\frac{20(K+11)({\cal J}_+'{\cal J}_3)}{3(K-4)^2}-
\frac{5(K+6))({\cal U}'{\cal J}_+)}{(K-4)^2}+
\frac{(K^2+7K+56){\cal J}_+''}{3(K-4)^2}, \nn  \\
W & = & {\cal G}^+_1 -\frac{100({\cal J}_+{\cal J}_-{\cal J}_3)}{(K-4)^2}+
\frac{10({\cal J}_+{\cal J}_-')}{3(K-4)}+
\frac{100(61K^2+327K+416)({\cal J}_3{\cal J}_3{\cal J}_3)}{9(K-4)^2(3K+8)^2}+
\nn \\
    & & \frac{40(K+1)({\cal T}{\cal J}_3)}{3(K-4)(3K+8)}+
\frac{10({\cal T}{\cal U})}{3(3K+8)}-
\frac{100({\cal U}{\cal J}_+{\cal J}_-)}{3(K-4)^2}+
\frac{125(K+4)(11K+16)({\cal U}{\cal J}_3{\cal J}_3)}{3(K-4)^2(3K+8)^2}
    + \nn \\
    & &
\frac{250K(K+6)({\cal U}{\cal U}{\cal J}_3)}{3(K-4)^2(3K+8)^2}+
\frac{25(K+16)({\cal U}{\cal U}{\cal U})}{9(K-4)(3K+8)^2}-
\frac{5(7K+32)({\cal U}{\cal J}_3')}{6(K-4)(3K+8)}+
 \nn \\
    & & \frac{20(K+6)({\cal J}_+'{\cal J}_-)}{3(K-4)^2}-
\frac{5(31K^2+32K-224)({\cal J}_3'{\cal J}_3)}{6(K-4)^2(3K+8)}-
\frac{5(7K^2+64K+32)({\cal U}'{\cal J}_3)}{6(K-4)^2(3K+8)}-\nn\\
    & &
\frac{5(K+16)({\cal U}'{\cal U})}{6(K-4)(3K+8)}-\frac{{\cal T}'}{6}+
\frac{(K+1)(3K-52){\cal J}_3''}{9(K-4)(3K+8)} +
\frac{(3K^2-4K+368){\cal U}''}{36(K-4)(3K+8)}. \label{realiz2}
\end{eqnarray}

The formulas \p{realiz2} give us the
desired realizations of the $W(sl(3|1),sl(3))$
superalgebra in terms of the currents of the linear superalgebra \p{superlin}.
For the values $K=-28/3,-1,52/3$ this realization is {\it the realization
of $W_3$ modulo null fields}.

\setcounter{equation}0
\section{Conclusion and outlook}

In the present Letter we construct explicitly the nonlinear algebras
$W(sl(4),sl(3))$ and $W(sl(3|1), sl(3))$ and find their realizations
in terms of currents spanning the corresponding
linearizing conformal algebras.
The specific structure of these algebras allows us to
construct modulo null fields realizations of the $W_3$ algebra that lies in
the cosets $W(sl(4),sl(3))/u(1)$ and $W(sl(3|1),sl(3))/u(1)$.
Such realizations exist for the following values of the
$W_3$ algebra central charge: $c_W=-30,-40/7,-98/5,-2$. The first two
values are listed for the first time, whereas for the last two values we
get the new realizations in terms of an
arbitrary stress tensor and $u(1)\times sl(2)$ affine currents.

Let us finish by presenting a conjecture about the spectrum of
central charges for the minimal models of the
considered $W(sl(4),sl(3))$ and $W(sl(3|1),sl(3))$ algebras.
The idea can be described as follows:
it is evident that for both linear algebras \p{WnewBosLin} and
\p{superlin}, after putting the null currents ${\cal G}^+_1$ and
${\cal G}^+_2$ to zero,
one can construct the stress tensor $T_d(z)$ which commutes with
the remaining currents. The central charge $c_d$ of this stress tensor
$T_d(z)$ is still connected with the central charges of the nonlinear
$W(sl(4),sl(3))$ and $W(sl(3|1),sl(3))$ algebras and the latter
can be expressed in terms of $c_d$. This is the general situation for the
conformal linearization procedure \cite{{sk1},{sk2},{sk3}}.
Moreover, in all known cases of linearizing algebras, the minimal models
for the Virasoro algebra spanned by $T_d(z)$ reproduce the minimal models
for the nonlinear algebras. One can assume that the same is true for the
$W(sl(4),sl(3))$ and $W(sl(3|1),sl(3))$ algebras.
If this conjecture is correct, then the central charge $c_d$
for the Virasoro minimal models
$$
c_d=1-6\frac{(p-q)^2}{pq}
$$
will induce the following values for the central charge of the
minimal models:
$$
c= \frac{(8p-15q)(4q-3p)}{pq} \quad \mbox{ for } W(sl(4),sl(3))\quad ,
$$
and
$$
c= \frac{(8p-3q)(2q-3p)}{pq} \quad \mbox{ for } W(sl(3|1),sl(3))\quad .
$$
Of course, this conjecture must be checked by the standard methods.

An interesting extension of the results presented here comes from
the consideration of the nonabelian case,
which corresponds to the algebras $W(sl(N+3),sl(3))$ and
$W(sl(3|N),sl(3))$.
For these algebras, the same procedure for constructing the
realizations of $W_3$ modulo null fields will give the series of the
central charges (with a manifest dependence on $N$). This work is
in progress.\vspace{0.5cm}\\

\noindent{\large\bf Acknowledgments}

We are grateful  to E. Bergshoeff, E. Ivanov, V. Ogievetsky and  M. Vasiliev
for many useful discussions. Two of us (S.K. and A.S.) acknowledge
the hospitality at INFN-LNF when this work was undertaken.
S.B. wishes to thank JINR and especially Rita and Victor Ogievetsky
for their kind hospitality when this work was completed.

This investigation has been supported in part by a grant of the
INFN, Fondo Scambi Internazionali, the
Russian Foundation of Fundamental Research,
grant 93-02-03821, and the International Science
Foundation, grant M9T300 and INTAS, grant 93-127.


\begin{thebibliography}{99}
\bibitem{a1} P. Bouwknegt and K. Schoutens, Phys. Rep. 223 (1993) 183;\\
L. F\'eher, L. O'Raifeartaigh, P. Ruelle, I. Tsutsui and A. Wipf,
     Phys. Rep. 222 (1992) 1.
\bibitem{sk1} S. Krivonos and A. Sorin, Phys. Lett. B335 (1994) 45.
\bibitem{sk2} S. Bellucci, S. Krivonos and A. Sorin, Phys. Lett.
                 B347 (1995) 260.
\bibitem{sk3} S. Krivonos and A. Sorin, {\it "More on the linearization
         of W algebras"}, JINR E2-95-151, hep-th/9503118.
\bibitem{rm} J.O. Madsen and E. Ragoucy, {\it " Secondary hamiltonian
         reductions"}, ENSLAPP-A-507-95, hep-th/9503042.
\bibitem{a2} A.B. Zamolodchikov and V.A. Fateev, Nucl. Phys. B280 (1987) 644.
\bibitem{a4} F.J. Narganes-Quijano, Int. J. Mod. Phys. A6 (1991) 2611.
\bibitem{a3} C.M. Hull and L. Palacios, Mod. Phys. Lett. A7 (1992) 2619.
\bibitem{a7} I. Bakas and E. Kiritsis, Int. J. Mod. Phys. A7 (Suppl. 1A)
(1992) 55.
\bibitem{a5} E. Bergshoeff, H.J. Boonstra and M. de Roo, Phys. Lett.
                   B292 (1992) 307.
\bibitem{a6} R. Blumenhagen, W. Eholzer, A. Honecker, K. Hornfleck
          and R. Hubel, Int. J. Mod. Phys. A10 (1995) 2367.
\bibitem{a8} F.A. Bais, T. Tjin and P. van Driel, Nucl. Phys. B357 (1991) 632.
\end{thebibliography}
\end{document}